\documentclass[prl,
reprint,aps,showpacs,amsmath,amssymb,showkeys,
nofootinbib,floatfix,superscriptaddress]{revtex4-2}
\usepackage[english]{babel}
\usepackage{graphicx}
\usepackage{color}
\usepackage{here}
\usepackage[dvipsnames]{xcolor}
\newcommand{\REV}[1]{{\color{black}{#1}}}
\newcommand{\FIN}[1]{{\color{black}{#1}}}

\begin{document}

\preprint{APS/123-QED}

\title{Collapse of Coherent Large Scale Flow in Strongly Turbulent Liquid Metal Convection}

\author{Felix Schindler}
\affiliation{Helmholtz-Zentrum Dresden-Rossendorf, 01328 Dresden, Germany}
\author{Sven Eckert}
\affiliation{Helmholtz-Zentrum Dresden-Rossendorf, 01328 Dresden, Germany}
\author{Till Zürner}
\affiliation{Technische Universität Ilmenau, 98684 Ilmenau, Germany}
\affiliation{UME, ENSTA Paris, Institut Polytechnique de Paris, 91120 Palaiseau, France}
\author{J\"org Schumacher}
\affiliation{Technische Universität Ilmenau, 98684 Ilmenau, Germany}
\author{Tobias Vogt}
\affiliation{Helmholtz-Zentrum Dresden-Rossendorf, 01328 Dresden, Germany}


\date{\today}

\begin{abstract}
The large-scale flow structure and the turbulent transfer of heat and momentum are directly measured in highly turbulent liquid metal convection experiments  for Rayleigh numbers varied between $4 \times 10^5$ and $\leq 5 \times 10^9$ and Prandtl numbers of $0.025~\leq~Pr~\leq ~0.033$. Our measurements are performed in two cylindrical samples of aspect ratios $\Gamma =$ diameter/height $= 0.5$ and 1 filled with the eutectic alloy GaInSn. The reconstruction of the three-dimensional flow pattern by 17 ultrasound Doppler velocimetry sensors detecting the velocity profiles along their beamlines in different planes reveals a clear breakdown \FIN{of coherence} of the large-scale circulation for $\Gamma = 0.5$.  As a consequence, the scaling laws for heat and momentum transfer inherit a dependence on the aspect ratio. We show that this breakdown of coherence is accompanied with a reduction of the Reynolds number $Re$. The scaling exponent $\beta$ of the power law $Nu\propto Ra^{\beta}$ crosses \FIN{eventually} over from $\beta=0.221$ to 0.124 when the liquid metal flow at $\Gamma=0.5$ reaches $Ra\gtrsim 2\times 10^8$ \FIN{and the coherent large-scale flow is completely collapsed}.
\end{abstract}

\maketitle
\begin{figure}
\includegraphics[width=\columnwidth]{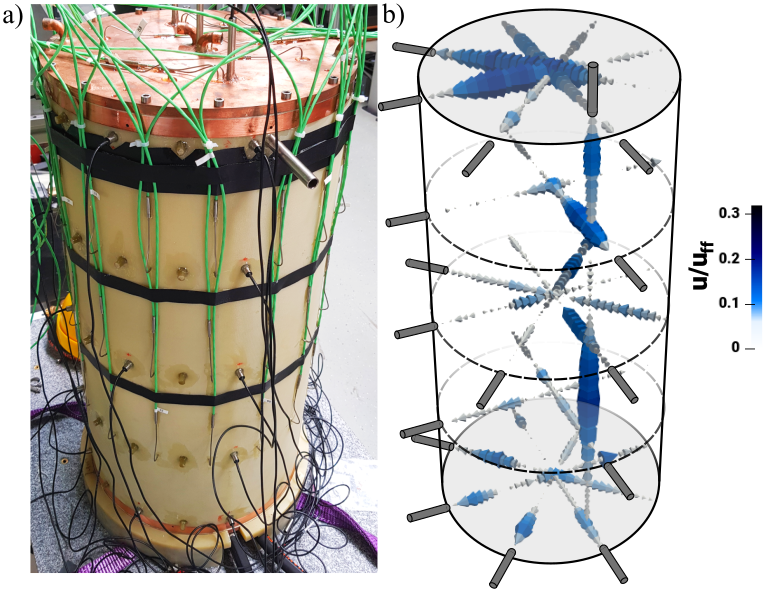}
\caption{Experimental setup. a) Photograph of the experimental setup without thermal insulation. b) Instantaneous large-scale flow visualization by means of UDV (grey sensors) in the measurement at $Ra=5\times 10^{9}$. The velocity magnitude $u$ is given in units of the free-fall velocity $u_{f\!f}=(g\alpha \Delta T H)^{1/2}$.}
\label{fig1}
\end{figure}
\begin{figure}
\includegraphics[width=0.9\columnwidth]{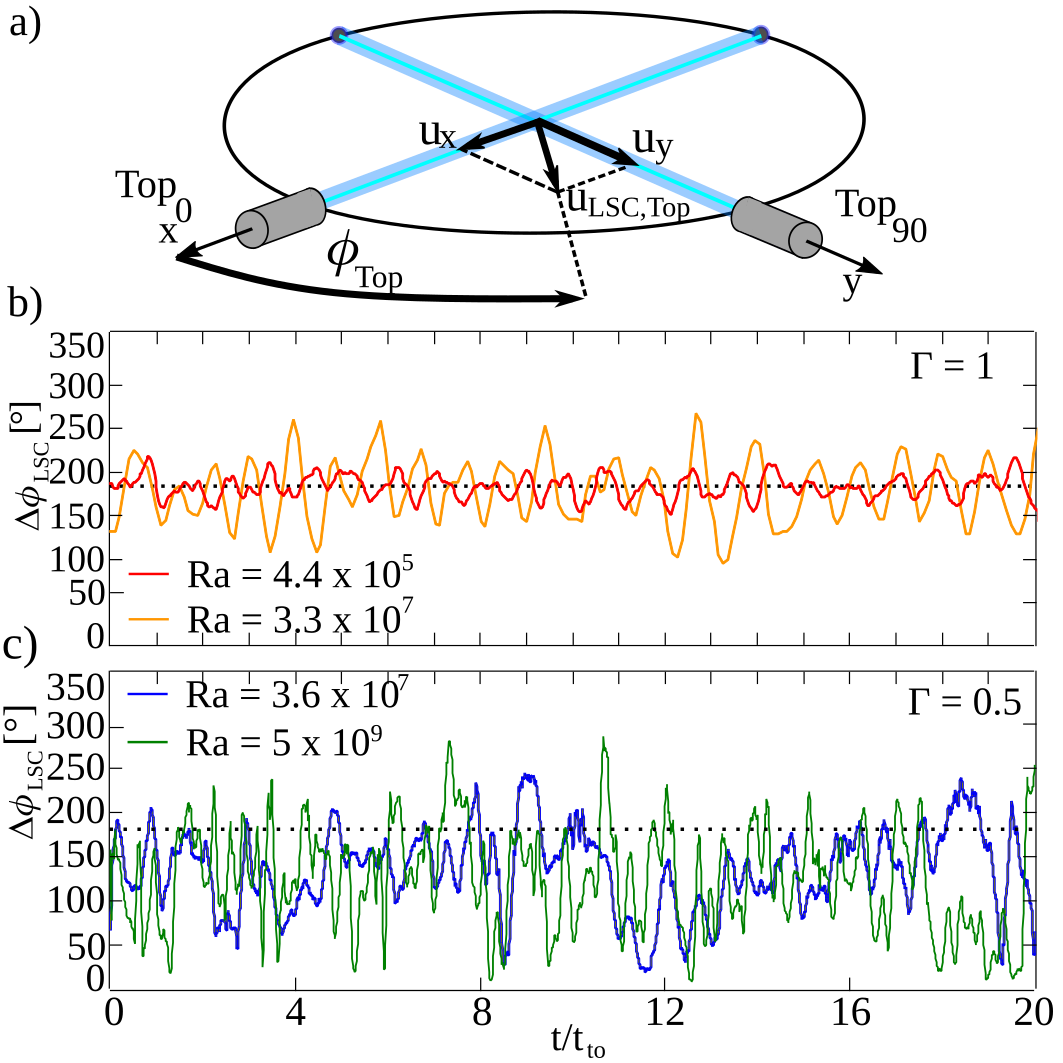}
\caption{Large-scale flow analysis. a) Determination of the LSC orientation using two crossed UDV sensors (see \cite{zuerner2019} for more details). b) Angular difference \FIN{$\Delta\phi_{LSC}\in[0^\circ,360^\circ[$} of the flow direction at the top and bottom plate vs. time normalized by the turnover time $t_{to}$ for two $Ra$ in the $\Gamma=1$ cell showing oscillations around a stable mean value of about 180$^\circ$ due to torsion and sloshing of a SR-LSC. c) Respective data for the $\Gamma=0.5$ cell revealing non-periodic frequent changes between various flow states.}
\label{fig2}
\end{figure}
\begin{figure}
\includegraphics[width=0.9\columnwidth]{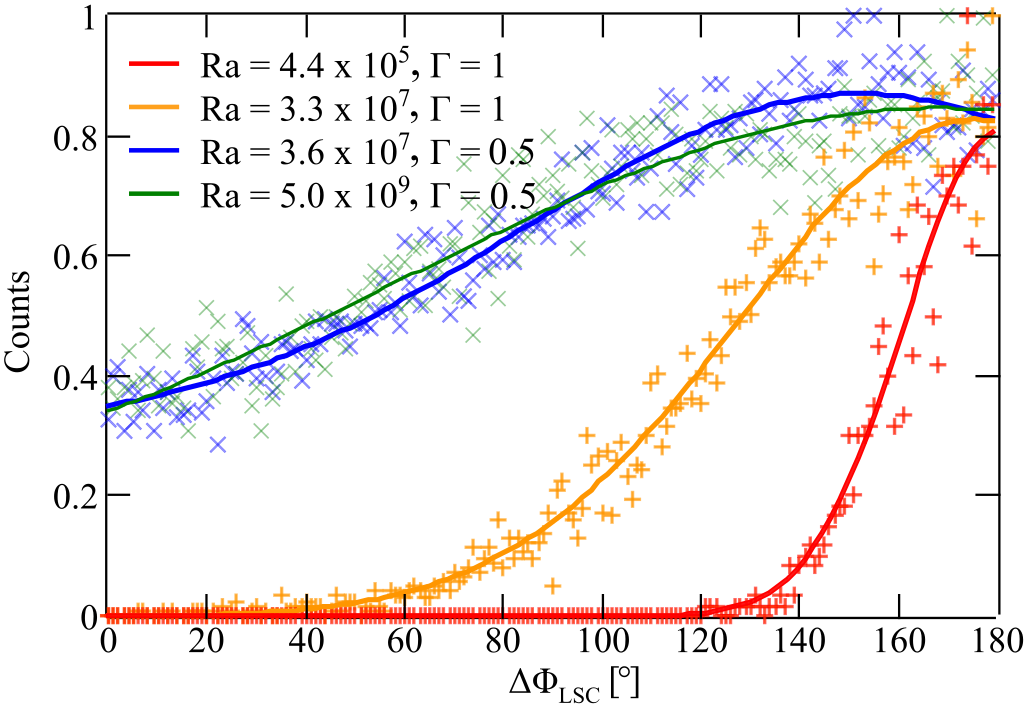}
\caption{Histograms of the angular difference $\Delta\Phi_\mathit{LSC}\in[0^\circ,180^\circ]$ for all measurements carried out at the same Rayleigh numbers that are exemplary shown in Fig. \ref{fig2}. The histograms are normalized by their respective maximal value. \FIN{The solid lines represent Gaussian fits.}}
\label{fig3}
\end{figure}
\begin{figure}
\includegraphics[width=0.9\columnwidth]{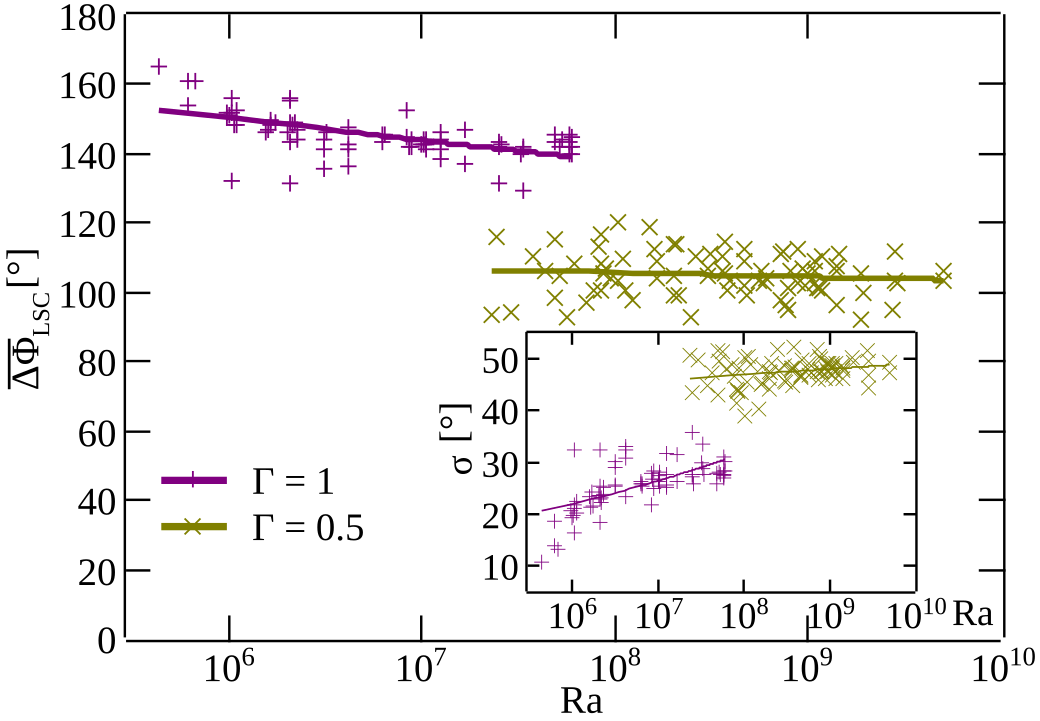}
\caption{Averaged values of the \FIN{$\Delta\Phi_{LSC}$} histograms vs. $Ra$. 
Inset: Corresponding standard deviation $\sigma$. 
}
\label{fig4}
\end{figure}

In turbulent convection flows, heat and momentum are primarily transported by a large-scale circulation (LSC) which is \FIN{built} up by a successive clustering of thermal plumes, fragments of the thermal boundary layer that rise from the top and bottom plates into the interior \cite{kadanoff2001, ahlers2009a, chilla2012}. The LSC is manifested as a single, partly twisted roll (SR) that fills a closed cuboid or cylindrical cell in case of width-to-height aspect ratios \REV{$\Gamma \approx 1$} \cite{xi2008, shi2012, brown2016, xu2021};  it appears as a whole coherent pattern of circulation rolls in convection layers with very large \REV{$\Gamma$} \cite{vonhardenberg2008, stevens2018, pandey2018, fonda2019,fodor2019,green2020,krug2020}. Even though the basic SR is superposed by three-dimensional dynamics of fluctuations, a number of investigations demonstrated that their coherence and role as a backbone of the heat transport remains intact, even for higher Rayleigh number $Ra$, a dimensionless measure of the vigor of convective turbulence, see e.g. refs. \cite{ahlers2009a,chilla2012,ahlers2009b,tsuji2005,vanderpoel2011,zuerner2019}. For \REV{$\Gamma < 1$}, the LSC forms multiple rolls arranged on top of each other \cite{xi2008, zwirner2018, zwirner2020a, zwirner2020b}. The particular LSC configuration determines the magnitude of transferred heat which is quantified by the Nusselt number $Nu$ \cite{bailoncuba2010, vanderpoel2011, zwirner2020b}. Furthermore, most theories of turbulent heat transfer \cite{shraiman1990, grossmann2000} rely on the existence of a mean wind, another notion for the LSC, that provides the major fraction of kinetic energy dissipation close to the plates and allows to separate this region from the interior. Most of the experimental studies are conducted in air or water, the simulations are typically run for Prandtl numbers $0.1 \leq Pr \leq 7$. It is thus still open how the LSC is connected to the turbulent transfer when we study convection beyond this parameter range, such as for very low $Pr$.\\
Our presented laboratory experiments in liquid metal convection extend this range to very low $Pr$. Here, we demonstrate the collapse of the LSC into a highly turbulent flow which causes a dramatic decrease of the amount of heat that is transferred across the fluid layer. Our experiment is \REV{designed for simultaneous measurement of temperature and velocity fields (see Fig.~\ref{fig1}(a)) which allows an unprecedented structural analysis of the LSC in liquid metals}. A total of 17 ultrasound Doppler velocimetry (UDV) sensors provide a volumetric reconstruction of the LSC in the opaque liquid metal flow to relate its structure to the turbulent heat transfer. Our data reflect a strong dependence of the turbulent heat transfer on the cell geometry; it also demonstrates that the transport mechanisms in convection can be altered profoundly when the Prandtl number of the fluid is taken to the lower limits that are possible in laboratory flows, all this in a parameter range that is inaccessible to long-term direct numerical simulations \cite{scheel2017}. We show that the observed LSC collapse causes a significantly smaller scaling exponent $\beta$ of the global heat transfer law $Nu\propto Ra^{\beta}$ that goes even below the value $\beta=1/4$ of an asymptotic two-dimensional theory \cite{busse1981}.\\
{\em Experimental setup.} \REV{Our study is performed in two upright cylindrical vessels with aspect ratios $\Gamma=D/H=1$ ($D=H=180$~mm) and $\Gamma=0.5$ ($H=2D=640$~mm) between two copper plates at constant temperatures and a thermally insulated sidewall.} The vessels are filled with the eutectic alloy GaInSn, which is liquid at room temperature. The flow is actuated by the temperature difference $\Delta T = T_{Bot}-T_{Top}$ between the heated bottom and the cooled top. The thermal driving is represented by the Rayleigh number $Ra = \alpha g H^{3}\Delta T/\nu\kappa$ and the Prandtl number is given by $Pr = \nu/\kappa$ with kinematic viscosity $\nu$, thermal diffusivity $\kappa$, thermal expansion coefficient $\alpha$, acceleration due to gravity $g$, cell diameter $D$, and height $H$. Respective thermophysical properties are found in \cite{plevachuk2014}. We consider $Ra$ ranges $4\times 10^{5} \leq Ra \leq 6 \times 10^{7}$ for $\Gamma=1$ and $2 \times 10^{7} \leq Ra \leq 5 \times 10^{9}$ for $\Gamma=0.5$.
\REV{Details of the experimental setup such as the sensor arrangement, $Nu$ measurement principle and tables with the measurement data can be found in the supplementary information \cite{supplement}.}
The high thermal conductivity of liquid metals requires special attention in the design of the thermal conditions at the plates. An assessment in this respect can be made by the Biot number \cite{oezisik1980,verzicco2004} which incorporates the effects of turbulent transport using the Nusselt number and is given by
\begin{equation}
Bi = Nu \left(\frac{\lambda}{\lambda_{C\!u}}\right)\left(\frac{H_{C\!u}}{H}\right)\,, 
\end{equation}
where $\lambda$ and $\lambda_{C\!u}$ are the thermal conductivity of the liquid metal and copper, respectively, and $H_{C\!u}$ denotes the thickness of the copper plates which have been chosen sufficiently thin to minimize thermal inertia. The boundary conditions are assumed to be isothermal for $Bi \ll 1$ \cite{xu2021b}. For our experiments, the Biot-number ranges between $0.03 \leq Bi \leq 0.125$ for $\Gamma=1$ and $0.01 \leq Bi \leq 0.019$ for $\Gamma=0.5$, respectively. Heat losses through the side walls and the copper plates are minimized by insulating the experiment with closed-cell polyethylene foam. \REV{We thus conclude that the thermal boundary conditions in our experiment satisfy the isothermal and adiabatic properties at the top/bottom plates and the sidewall, respectively \cite{supplement}.}\\  
Linear velocity profiles are measured by UDV \cite{takeda2002, eckert2002} with a total of 17 sensors, 16 of which are located at different azimuthal positions to determine the radial velocity distribution in five different horizontal planes (see Fig.~\ref{fig1}(b)). The remaining sensor measures the vertical velocity component along the cell height at $r/R=2/3$ with the cell radius $R=D/2$. When all ultrasonic sensors are scanned in series, a sampling rate of 1 Hz and an accuracy of $0.1$ mm/s are achieved.\\
{\em Large-scale flow structure.} Figure \ref{fig1}(b) shows a snapshot of a typical incoherent flow pattern at $Ra=5\times 10^9$ which is interrupted rarely by short-term periods with a single roll structure. The supplemental movie \cite{supplement} displays an example for the visualization of the LSC structure by simultaneous temperature and velocity measurements in the cell at $\Gamma=0.5$ and $Ra = 5 \times 10^9$. Already at first glance, the volatile character of the flow becomes obvious. Drastic changes can be observed on short time scales involving frequent reorientations as well as rapid and irregular rotations. This accumulates to a disintegration of the SR structure and temporary transitions into double- or multi-roll structures. The coherence level of the LSC is apparently quite low.\\
A robust criterion to estimate the coherence of the LSC is to determine the phase correlation of the mean flow direction at both copper plates. We rely on two radial UDV sensors, placed at $90^{\circ}$ to each other, which determine the mean flow direction \FIN{$\phi_{Top}$} and \FIN{$\phi_{Bot}$} as well as the time-averaged LSC velocity magnitude $u_{LSC}=\overline{(u_{LSC,Top}+u_{LSC,Bot})/2}$ near the viscous boundary layers at a distance of 10 mm from the top/bottom plate, as shown in Fig.~\ref{fig2}(a) (see~\cite{zuerner2019}~for~more~details). From the angular difference \FIN{$\Delta\phi_{LSC}=|\phi_{Top}-\phi_{Bot}|$} between the top and bottom flow, the coherence of the flow can be assessed in a relatively simple way \cite{xi2008}. In the case of a SR-LSC, the fluid at the top and the bottom is expected to move in opposite directions, causing an angular difference \FIN{$\Delta\phi_{LSC}\sim 180^\circ$}, while the formation of a double-roll structure is associated with a vanishing angular difference. From \cite{ahlers2009a,xi2008}, the well-proven approach is known to evaluate the coherence of the LSC based on the temperature measurements at three different heights, where the position of upward (downward) flow was determined by locating the maximum (minimum) of the temperature distribution on the circumference. The main flow direction can be accessed directly by UDV and has not to rely on  an indirect determination via temperature, which is affected by the large thermal diffusivity at low $Pr$.\\
%
In Fig. \ref{fig2}, the measured angular difference is plotted versus \REV{time normalized by the turnover time $t_{to} = L/u_{LSC}$} assuming a SR-LSC, where the LSC path length $L$ is taken as the largest ellipse that fits into the vertical mid-plane of the convection cell. Fig. \ref{fig2}(b) contains the results for the $\Gamma=1$ cell at two different $Ra$. The stable average value of \FIN{about} $180^\circ$ indicates a coherent SR-LSC. The regular oscillations, which are visible particularly at the higher $Ra$ and correlate with the turnover time, provide a clear indication of the torsional and sloshing modes typically observed in $\Gamma=1$ cells \cite{brown2009, zuerner2019}. The corresponding data for $\Gamma=0.5$ in Fig.~\ref{fig2}(c) display rapid fluctuations and weak correlation between the flow at both plates. The mean value deviates significantly from $180^\circ$. However, the SR state is not just replaced by another LSC structure, e.g. a double-roll, since \FIN{$\Delta\phi_{LSC}$} does not level off to a value close to zero.\\
In addition to the results in Fig.~\ref{fig2}, we plot in Fig.~\ref{fig3} corresponding histograms of $\Delta\Phi_{LSC} \in[0^\circ,180^\circ]$, for which the values larger $180^\circ$ are transformed \FIN{from $\Delta\phi_{LSC}$} by \FIN{$\Delta\Phi_{LSC}=360^\circ-\Delta\phi_{LSC}$} to fit into the domain. In this plot, a coherent single-roll LSC would exhibit a narrow peak \FIN{near} $180^\circ$. The measurements show that the flow increasingly deviates from the SR state as $Ra$ increases and $\Gamma$ decreases. For the measurements at $\Gamma=0.5$, there is even a trend towards a uniform distribution of the angular differences. In such a case the flows at the top and bottom would be fully uncorrelated;
an inherent coupling between the flows at bottom and top (which exists in case of $\Gamma=1$) no longer occurs here.\\
Mean value \FIN{$\overline{\Delta\Phi}_{LSC}$} and standard deviation \FIN{$\sigma$ of $\Delta\Phi_{LSC}$} are drawn in Fig. \ref{fig4} versus $Ra$. For $\Gamma=1$ and low $Ra$, the mean value is quite close to the ideal value of $180^\circ$ for the SR-LSC and the standard deviation is comparatively small. With increasing $Ra$, we find a decrease in the mean value and a growing standard deviation. This demonstrates that the flow becomes more turbulent and, conversely, the coherence decreases. \FIN{Just changing the aspect ratio to $\Gamma = 0.5$ reduces the mean angle difference and increases the standard deviation drastically.} A further increase of $Ra$ does not \REV{show} a significant effect.

\begin{figure}
\includegraphics[width=0.9\columnwidth]{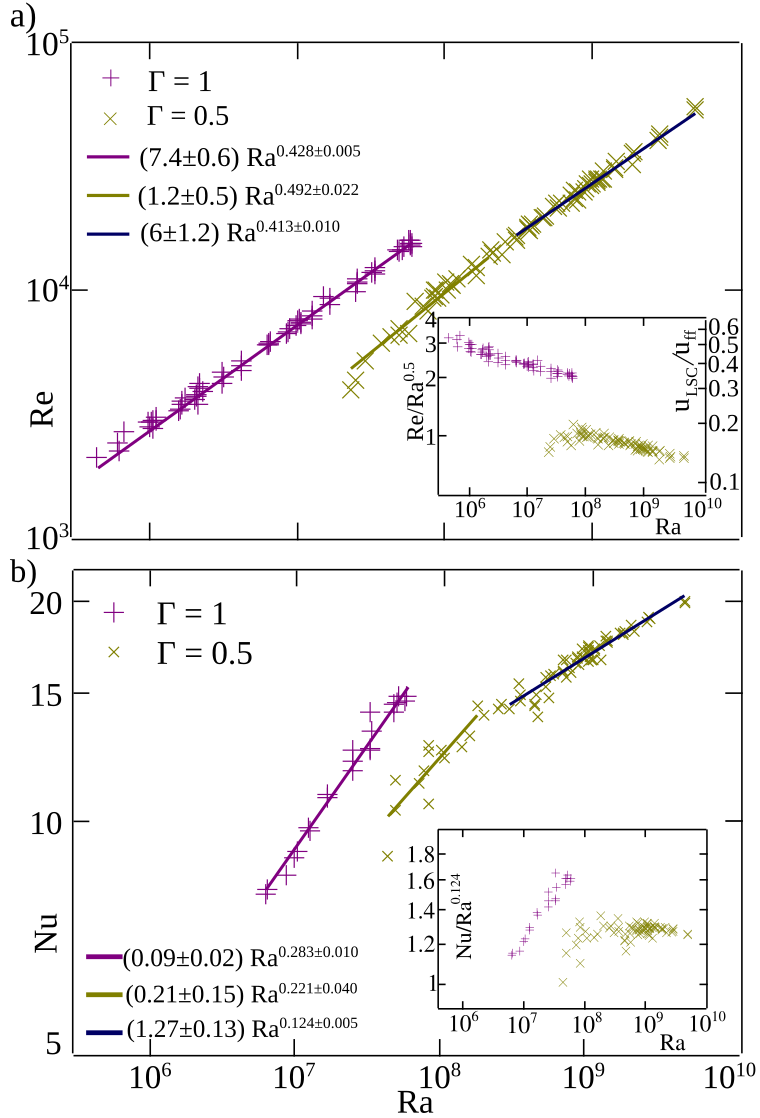}
\caption{Global transport laws ($Pr = 0.029$). a) Reynolds number $Re$ versus Rayleigh number $Ra$. Inset: Compensated plot of $Re/Ra^{1/2}$ and corresponding $u_{LSC}/u_{f\!f}$ versus $Ra$. b) Nusselt number $Nu$ versus $Ra$. Inset: \REV{Compensated plot $Nu/Ra^{0.124}$ versus $Ra$}.
\label{fig5}}
\end{figure}

{\em Scalings and transport laws.} Previous liquid metal convection experiments in $\Gamma \geq 1$ and moderate $Ra$ exhibit a robust and coherent LSC \cite{vogt2018,zuerner2019}. This is associated with intense flow velocities of $u_\mathit{LSC}/u_{f\!f} = (Ra/Pr)^{-0.5}Re \approx 0.7$ in a $\Gamma=2$ cylinder \cite{vogt2018} and a $\Gamma=5$ rectangular box \cite{vogt2021}, where \FIN{$Re=u_{LSC}H/\nu$ is the Reynolds number} and the free-fall velocity $u_{f\!f} =\sqrt{g\alpha\Delta T H}$ is the theoretical upper velocity limit in case the potential buoyancy is completely converted into momentum. In contrast, our measurements performed in $\Gamma=0.5$ reveal an almost complete breakdown of the SR-LSC \FIN{for the whole $Ra$ range}. This is associated with a drastic reduction of the LSC velocity as reflected in a ratio of $u_{LSC}/u_{f\!f} < 0.2$ (see inset of Fig.~ \ref{fig5}(a)). The SR-LSC is still clearly pronounced in $\Gamma=1$ cylinders \cite{zuerner2019}, although our velocity measurements there already show a \FIN{smaller} ratio $u_{LSC}/u_{f\!f} \approx 0.4$. The corresponding $Re$ is plotted for both aspect ratios in Fig.~\ref{fig5}(a). For the same $Ra$, significantly higher $Re$ values occur in the $\Gamma=1$ cell with a scaling of $Re \varpropto Ra^{0.428}$. \REV{The data at $\Gamma=0.5$ indicate a gradual change of the scaling of $Re(Ra)$ at $Ra\approx 2 \times 10^8$.}\\
The Nusselt number measurements in Fig. \ref{fig5}(b) give a qualitatively similar picture\REV{, although the accessible measurement range of $Nu$ is narrower than that of the UDV measurements, which reliably measure even the lowest velocities of 0.5 mm/s.} The data is referred to $Pr = 0.029$ where the deviations in the Prandtl number are accounted for by assuming a $Pr$ dependence of the heat transfer of $Nu\varpropto Pr^{0.14}$ and correcting the $Nu$ values accordingly. This power law is based on simulations by Verzicco and Camussi \cite{verzicco1999} which provide data for $Pr <$ 1 \cite{ahlers2009b, stevens2013}. The measured scaling law $Nu\varpropto Ra^{0.28\pm 0.01}$ for $\Gamma=1$ is in good agreement with existing data at similar $Pr$ \cite{cioni1997,glazier1999,zuerner2019}. In contrast, the $Nu$ values and the scaling exponent $\beta$ for $\Gamma=0.5$ are lower than those for $\Gamma=1$ at comparable $Ra$. \REV{Moreover, our data indicate two regions with different power laws}: (i) for $4\times 10^7 \le Ra \lesssim 2\times 10^8$ we find $Nu\varpropto Ra^{0.22\pm 0.04}$ and (ii) for $2\times 10^8 \lesssim Ra \le 5\times 10^9$ the scaling $Nu\varpropto Ra^{0.124\pm 0.005}$ results from the data. \FIN{The transition between the two scalings is observed at approximately the same $Ra$ at which the change in $Re$ scaling is found.}\\
%
{\em Final discussion.} Our heat transfer measurements differ from those at $Pr\sim 1$ \cite{ahlers2012,he2012,ching2016} where no significant differences were found for the heat transfer scaling for $\Gamma=0.5,\,1$. The same holds for a comparison of $\Gamma=0.1, 1$  in direct numerical simulations for $10^8\le Ra\le 10^{10}$ \cite{Iyer2020}. Our findings differ also from measurements in mercury at $Pr \thickapprox$ 0.025 \cite{naert1997,cioni1997,glazier1999}. Naert et al. \cite{naert1997} obtained exponents of $\beta$ = 0.27, 0.25 and 0.28 for aspect ratios of $\Gamma$ = 2, 1, and 0.5, respectively. Glazier et al. \cite{glazier1999} reported a scaling $Nu \varpropto Ra^{0.29}$ for the entire range $10^5 < Ra <10^{11}$ using cells with $\Gamma$ = 0.5, 1, and 2. Cioni et al. \cite{cioni1997} identified three different scalings at $\Gamma=1$ between $Ra = 7\times10^6$ and $2\times10^9$. Interestingly, their scaling law for the heat transfer changes from $Nu \varpropto Ra^{0.26}$ to $Nu \varpropto Ra^{0.20}$ which occurs \FIN{around} $Ra = 4.5 \times10^8$, i.e., close to our value of $Ra \approx 2 \times10^8$. The scaling for $Ra < 4.5\times 10^8$ agrees well with our results in $\Gamma$=1. It is also worth noting that our $\Gamma$ = 0.5 measurements for $Ra \lesssim 2\times 10^8$ show a comparable scaling as found by Cioni et al. in their region II \cite{cioni1997}. Both experiments \cite{cioni1997,glazier1999} date back more than 20 years and do not provide any direct velocity measurements to characterize the LSC and its possible breakdown. Their analysis was based on temperature measurements only. We recall also that the $Nu$ measurements of both works come to different results. A comparative analysis of our data with both pioneering works is hardly possible. Furthermore, we note that the exponent $\beta=0.22$ for $Ra \leq 2\times10^8$ is in agreement to predictions from the Grossmann-Lohse theory for low Prandtl numbers \cite{grossmann2000}. The exponent of $\beta=0.124$ for $Ra \geq 2\times10^8$, however, falls even below a prediction of $\beta=1/4$ from a two-dimensional asymptotic theory \cite{busse1981} Thus, the $Nu(Ra)$ scaling law in $\Gamma$ = 0.5 for $Ra \gtrsim 2\times10^8$ reveals a very small $\beta$, which to the best of our knowledge, has not been reported anywhere before. 

\FIN{The change in scaling at $Ra \sim 2\times 10^8$ might originate from the transition of a highly fluctuating, partially decoherent to a fully collapsed large-scale flow.}
A similar sharp decrease of the $Nu(Ra)$ scaling with increasing $Ra$ is known only from an experimental study in a cylindrical gap \cite{xie2018} where a sharp transition from $Nu \varpropto Ra^{0.274}$ to $Nu \varpropto Ra^{0.17}$ is found at $Ra \thickapprox 6.35 \times 10^8$. The authors trace this transition to a change from a high-symmetry, coherent state to a low-symmetry, turbulent state. The present low-$Pr$ measurements consistently show a simillar loss of coherence in the flow at \FIN{$\Gamma=0.5$}, which leads to significantly lower velocity amplitudes and reduced heat transport in comparison to the \FIN{stable} LSC at $\Gamma=1$. This explanation is supported by the fact that the effect works in the opposite direction as well. In \cite{chong2017} and \cite{vogt2021} it was shown that an increase of flow structure coherence can increase the heat transport. It is also mentioned that higher $Pr$ seem to cause more stable LSC configurations \cite{ciliberto1996,xia1997}. We conclude that our present knowledge regarding the LSC properties at low $Pr$ and large $Ra$ for $\Gamma < 1$ is still incomplete. Theoretical models assume a coherent wind. Its breakdown alters heat and momentum transport in ways that ask to our view for further detailed research on this topic.

\acknowledgments{This work is supported by the Deutsche Forschungsgemeinschaft with Grants No. VO 2332/1-1 and SCHU 1410/29-1.}

\end{document}